\documentclass[amsmath,preprint,prb,11pt,nobibnotes]{revtex4-1}

\usepackage{graphicx} 
\usepackage{bm}
\usepackage{hyperref} 
\hypersetup{breaklinks=true}

\begin{document}

\title{F{\"o}rster resonance energy transfer, absorption and emission 
spectra in multichromophoric systems: III. Exact stochastic path integral
evaluation}

\author{ Jeremy M. Moix } 
\author{ Jian Ma } 
\author{ Jianshu Cao }
\email{jianshu@mit.edu} 
\affiliation{ 
   Department of Chemistry, 
   Massachusetts Institute of Technology, 
   77 Massachusetts Avenue, 
   Cambridge, MA 02139 }

\date{\today}

\begin{abstract}

A numerically exact path integral treatment of the absorption and
emission spectra of open quantum systems is presented that requires only the
straightforward solution of a stochastic differential equation.
The approach converges rapidly enabling the calculation of spectra 
of large excitonic systems across the complete range of system
parameters and for arbitrary bath spectral densities.
With the numerically exact absorption and emission operators 
one can also immediately compute energy transfer rates using the 
multi-chromophoric F{\"o}rster resonant energy transfer formalism.
Benchmark calculations on the emission spectra of two level systems 
are presented demonstrating the efficacy of the stochastic approach.
This is followed by calculations of the energy transfer 
rates between two weakly coupled dimer systems as a function of 
temperature and system-bath coupling strength. 
It is shown that the recently developed hybrid cumulant expansion 
is the only perturbative method capable of generating uniformly
reliable energy transfer rates and spectra across a broad range of system 
parameters. 

\end{abstract}

\maketitle

\section{Introduction}

The far-field absorption and emission spectra are standard experimental tools
in the characterization of excitonic systems.  
The temperature and solvent dependence of these spectra are often
used to extract a wealth of information on, for example,
the microscopic geometry of the constituent chromophores, the
coupling strength between the excitonic system and its environment, as well as
the relative importance of heterogeneous broadening 
mechanisms.\cite{beljonne09,heijs05b,renger01} 
Despite this wide applicability, at present there are still relatively few
theoretical approaches that are capable of providing uniformly reliable 
estimates of the spectra of open quantum systems. 
Only in the limiting case that the spectrum arises from a single isolated
electronic transition can the exact absorption and emission spectrum be
obtained analytically (up to a numerical integration) through cumulant
expansion techniques.\cite{mukamel99}
However, in the more common setting, wherein the excitonic complexes are 
comprised of multiple coupled chromophores, then one must, in general, 
resort to numerical methods.
Unfortunately, there is no numerically exact approach currently available 
for systems containing more than a few chromophores that is valid 
over a large range of the parameter space.
As a result, comparisons between many interesting experimental systems 
and their corresponding microscopic theoretical models are often out of reach.
One of the central aims of this work is to fill this gap.
Here we present an efficient path integral approach that allows one
to compute the numerically exact absorption and emission spectra 
of multi-chromophoric open quantum systems.

Due to the lack of robust exact methods, one often turns to perturbative 
techniques.
As detailed in the preceding papers of this series
(henceforth referred to as papers I\cite{ma14a} and II\cite{ma14b})
many of the standard approximate approaches are capable of generating 
reliable absorption spectra, as only the short time dynamics of a factorized
system-environment initial state are required.
The emission spectrum, however, presents a much more challenging problem.
In this case, the real-time dynamics evolve from the correlated 
equilibrium state of the entire excitonic system and its environment.
Unless the system-bath coupling is very weak, perturbative
treatments often generate qualitatively incorrect emission spectra, and
generally become even worse as the temperature is lowered.\cite{ma14a,banchi13}
One of the major results of the previous papers in this series was a 
hybrid perturbative method capable of providing reliable emission spectrum 
over a broad range of system parameters.\cite{ma14b}
The approach is not fully perturbative in that it combines the knowledge of 
the numerically exact equilibrium reduced density matrix 
--which can be obtained relatively easily through imaginary time 
path integral methods--\cite{moix12}
with an approximate cumulant expansion of the remaining real time dynamics.
This initial state correction becomes essential at low temperatures
or strong coupling.
The hybrid cumulant expansion (HCE) thus greatly extends the parameter 
regimes accessible
to perturbative methods and generally improves the quality of the results.

In the context of numerically exact treatments of the emission spectrum,
such as the hierarchy equation of motion (HEOM) or quasi-adiabatic path
integral approach (QUAPI),\cite{Tanimura2006,Makri1995}
the problem of a correlated system-bath initial state is overcome by simply 
preparing a factorized state sufficiently far in the past such that
the system has reached equilibrium at time zero. 
As a result, these approaches require an initial lengthy propagation of 
the reduced density matrix to equilibrium before the dynamics of 
the dipole correlation function can be calculated.\cite{jing13}
Furthermore, as these approaches rely on efficient representations
of the influence functional, they are generally restricted to environments
that are not strongly non-Markovian.
The main result of this paper is a stochastic path integral approach 
that circumvents many of the restrictions imposed by other 
numerically exact methods, and in particular, is applicable for arbitrary
spectral densities and temperatures.

While an exact calculation of the absorption and emission spectra
is important in its own right, it also provides an additional benefit.  
That is, one can immediately compute energy transfer rates between 
weakly coupled excitonic systems using the multi-chromophoric F{\"o}rster
resonant energy transfer (MCFT) formalism. 
The MCFT framework is a generalization of the standard F{\"o}rster theory
to the situation where the donor or acceptor complex consists of 
multiple coupled chromophores,\cite{sumi99,jang04}
and has gained recent attention as this scenario appears to be one of most
common motifs employed in the highly efficient energy transfer networks
found in biological systems.
For example, the light harvesting systems found in both green and purple 
bacteria are comprised of independent complexes of strongly coupled 
chromophores that form the base units for large-scale energy transfer 
networks.\cite{sener07,olson04}

In Sec.~\ref{sec:formalism} the details of the MCFT formalism are presented.
There it becomes apparent that the key quantities necessary for computing
energy transfer rates are generalized operators related to the absorption 
spectrum of the acceptor complex and the emission spectrum of the donor.
Then we present the path integral treatment of the absorption
operator and demonstrate that it may be efficiently computed as the 
solution to a straightforward stochastic differential equation.
This approach is then generalized to the emission spectrum by taking
advantage of the detailed balance condition that relates the emission operator
to its corresponding absorption operator evolving in a complex time.
Following these formal developments, numerical calculations
are presented for model two level systems that can be reliably benchmarked 
against the HEOM approach.
The temperature dependence of the emission spectrum is presented, followed
by systematic calculations of the MCFT rate as a function of the 
temperature and system-bath coupling strength.
It is observed that the hybrid cumulant expansion technique
developed in paper II is the only perturbative approach
that provides uniformly reliable results for the energy transfer 
rates.\cite{ma14b}
In a forthcoming work, the path integral and HCE methods will be 
used to provide the first systematic analysis of 
the energy transfer rates between two B850 complexes in the 
light harvesting system LH2.\cite{moix14b}

\section{MCFT Formalism}\label{sec:formalism}

The MCFT formalism has been expounded in the previous papers in this series.
Here we provide only the salient details necessary to keep
the presentation self contained.
The total system is composed of a donor complex consisting of $N_{\rm D}$ 
chromophores that is weakly coupled to an acceptor complex of $N_{\rm A}$ 
chromophores.
The Hamiltonian for the entire donor-acceptor system is then
\begin{equation}
    H =  H^{\rm D} +  H^{\rm A} +  H^{\rm DA}\;,
\end{equation}
where $ H^{\rm D(A)}$ denotes the Hamiltonian operator of the 
multi-chromophoric donor (acceptor) complex along with its associated 
thermal environment. 
The excitonic coupling between the donor and acceptor systems
is characterized by $H^{\rm DA}$, which, within the local basis of the 
single-excitation subspace of the donor and acceptor, is given by 
\begin{equation}
   H^{\rm DA} =\sum_{n=1}^{N_{\rm D}}\sum_{m=1}^{N_{\rm A}} J_{nm}^{\rm DA}
               \left|D_n\right\rangle \left\langle A_m\right|
               \;.\label{eq:h_da}
\end{equation}
The Hamiltonian of an individual complex is modeled as
a general open quantum system,
\begin{equation}
   H^\alpha =  H_{\rm s}^{\alpha} +  H_{\rm b}^{\alpha} +  H_{\rm sb}^{\alpha}
    \,
\end{equation}
where the label $\alpha\in(D,A)$ serves to distinguish between the 
donor and acceptor systems.
The free excitonic Hamiltonian of each complex is given by
\begin{equation}
   H_{\rm s}^{\alpha} = \sum_{m=1}^{N_\alpha} 
              \left( \epsilon_m^{\alpha} + \lambda_m^\alpha \right)
              \left| \alpha_m \right \rangle\left\langle \alpha_m \right|
              + \sum_{n\neq m}^{N_\alpha} t_{nm}^\alpha 
              \left| \alpha_n \right \rangle\left\langle \alpha_m \right|
              \;,
\end{equation}
where $\epsilon_m$ is the excitation energy of the $m$-th chromophore,
$t_{nm}$ denotes the intra-complex electronic couplings
and $\lambda_m$ is the environment-induced reorganization energy.
The free bath Hamiltonian is
\begin{equation}
   H_{\rm b}^{\alpha}  = \sum_{m=1}^{N_\alpha} \sum_k 
   \hbar\omega_{m,k}^{\alpha} b_{m,k}^{\alpha\dagger} b_{m,k}^{\alpha}
   \;,
\end{equation}
where $b_{m,k}^{\alpha\dagger}$ ($b_{m,k}^\alpha$) denotes the
respective creation (annihilation) operator of the $k$-th mode of 
the bath with frequency $\omega_{m,k}^\alpha$, and
coupled to chromophore $m$ on the excitonic complex labeled by $\alpha$.
The system bath coupling is linear in the bath coordinates, 
and assumed to modulate only the excitation energies,
\begin{equation}
   H_{\rm sb}^{\alpha}  = \sum_{m=1}^{N_\alpha} V_m^\alpha \sum_k g_{m,k}^{\alpha}
   \left(b_{m,k}^{\alpha\dagger} + b_{m,k}^{\alpha}\right) \;,
\end{equation}
where 
$V_m^\alpha = \left| \alpha_m \right \rangle\left\langle \alpha_m \right|$ 
and $g_{m,k}^\alpha$ denotes the coupling strength.

Assuming that the exciton lifetime is much longer than the
timescale associated with the energy transfer, then relaxation to 
the ground state can be safely ignored,
and the population transfer rate between the donor and acceptor systems 
is given by the MCFT rate formula, 
\begin{equation}
   k = 2 {\rm Re} \int_0^\infty dt\; {\rm Tr}
         \left[ \left(H^{\rm DA}\right)^\top  \mathbf E^{\rm D}(t) 
             H^{\rm DA}  \mathbf I^{\rm A}(t)\right] \;,
   \label{eq:mcfret}
\end{equation}
which can be easily obtained from the golden rule expression
as shown in section II of paper I.\cite{ma14a}
The absorption operator of the acceptor, $\mathbf I^{\rm A}(t)$, 
and emission operator of the donor, $\mathbf E^{\rm D}(t)$,
appearing in Eq.~\ref{eq:mcfret} are formally defined as:
\begin{align}
    \mathbf I^{\rm A}(t) & = 
    {\rm Tr_b}\left[ e^{-\frac{i}{\hbar} H^{\rm A}t} \rho^{\rm A} 
      e^{+\frac{i}{\hbar}  H_{\rm b}^{\rm A} t}
      \right] \;,\label{eq:It} \\
    \mathbf E^{\rm D}(t) & = 
      {\rm Tr_b}\left[ e^{+\frac{i}{\hbar} H^{\rm D}t} \rho^{\rm D} 
      e^{-\frac{i}{\hbar}  H_{\rm b}^{\rm D} t}\right] 
      \;. \label{eq:Et}
\end{align}
In the case of the absorption operator, the initial density matrix 
corresponds to a factorized state of the system and bath, 
$\rho^{\rm A}=I_{\rm s}\otimes e^{-\beta H_{\rm b}^{\rm A}}
/Z_{\rm b}^{\rm A}$, due to the assumption of a Franck-Condon
transition from the ground state. 
The steady-state emission, however, occurs after the total system
has equilibrated within the single excitation manifold. 
Thus, the initial state in Eq.~\ref{eq:Et}
corresponds to the equilibrium state of the entire system and bath,
$\rho^{\rm D} = e^{-\beta H^{\rm D}}/Z^{\rm D}$, where 
$Z^{\rm D} = {\rm Tr}\left[e^{-\beta H^{\rm D}}\right]$ 
is the partition function of the donor.
As discussed extensively in the previous papers of the series, the 
difference in the initial states is the key feature
that distinguishes the absorption from the emission operator,
with the correlated initial condition in Eq.~\ref{eq:Et} 
leading to a substantially more involved calculation.

\subsection{Detailed Balance}

The absorption and emission spectra obey a well-known detailed balance
condition, and it is readily apparent that their corresponding operators in 
Eqns.~\ref{eq:It} and~\ref{eq:Et} must obey a similar relation.
In the frequency domain, the detailed balance condition for the
operators reads 
\begin{equation}
   \mathbf E^{\rm D}(\omega) =
      \frac{e^{\hbar \beta\omega}}{Z} \mathbf I^{\rm D}(\omega) 
   \;,\label{eq:db_w}
\end{equation}
where $Z= Z^{\rm D}/Z_{\rm b}^{\rm D}$.
Thus, in principle, knowledge of the absorption operator allows for 
a straightforward determination of the corresponding emission operator.
However, in practice, the thermal prefactor exponentially amplifies any 
error in the absorption data leading to an ill-conditioned numerical problem.
As a result, Eq.~\ref{eq:db_w} is generally of little practical use
outside of the very high temperature limit.

An alternative approach can be based on the observation that in the 
time domain, the detailed-balance condition takes the form
\begin{equation}
   \mathbf E^{\rm D}(t)^* = \frac{1}{Z} \mathbf I^{\rm D}(t-i\hbar\beta) 
   \;, \label{eq:db_t}
\end{equation}
where the asterisk denotes complex conjugation.
That is, through the straightforward substitution, 
$t\rightarrow t- i\hbar\beta$, 
the time evolution of the emission operator of the donor 
is equivalent to that of the absorption operator, 
except that the dynamics evolves in complex time rather than in purely 
real time.
In contrast to the frequency-domain detailed balance relation, the 
time-domain version in Eq.~\ref{eq:db_t} is free from numerical 
instabilities and forms the basis for the developments presented here.
Here we employ the path integral formalism to develop an
exact and efficient numerical treatment of the spectral operators
rather than pursue perturbative approaches as were explored in the 
previous papers of this series. 
In the following subsection, the stochastic path integral representation 
for the absorption operator is presented, and then generalized to the case of
emission through the rotation from real time to complex time suggested by 
Eq.~\ref{eq:db_t}.

\subsection{Absorption Operator}
\label{sec:abs}

As can be seen from Eq.~\ref{eq:It}, the absorption operator does not require 
the full time evolution of the reduced density matrix. 
The bath evolves both forward and backward in time,  
but the system is only propagated forward in time.
As a result, we can still take advantage of the influence functional formalism 
from the path integral approach to open quantum systems,\cite{grabert88}
but only require a single path for the system variables.
Thus the absorption operator can be determined from the path integral 
expression
\begin{equation}
   U^{\rm A}(t) = \int {\mathcal D}[\sigma] 
   e^{\frac{i}{\hbar} S_0^{\rm A}[\sigma]} F[\sigma]
   \label{eq:abs_pi}  \;,
\end{equation}
where $S_0^{\rm A}[\sigma]$ denotes the action associated with the free system 
Hamiltonian of the acceptor, $H_{\rm s}^{\rm A}$, and the standard boundary
conditions of the paths have been suppressed for clarity.
The Feynmann-Vernon influence functional, $F[\sigma]$, obtained
by integrating out each of the $N_{\rm A}$ independent baths is
given by
\begin{equation}
   F[\sigma] = \prod_{n=1}^{N_{\rm A}} 
      \exp\left(-\frac{1}{\hbar} \int_0^t dt' \int_0^{t'} d{t''} 
      V_n^{\rm A}(\sigma(t'))V_n^{\rm A}(\sigma(t'')) C_n(t'-t'') \right) 
      \;. \label{eq:if}
\end{equation}
All of the microscopic details of the baths that are relevant to the system
dynamics enter through their respective correlation functions in the 
influence functional, which take the standard form,
\begin{equation}
   C_n(t) = \frac{1}{\pi} \int_0^\infty d\omega\; 
     J_n(\omega)\left[\coth(\hbar\beta\omega/2)\cos(\omega t) 
        - i\sin(\omega t)\right]
     \;,\label{eq:ct}
\end{equation}
with the spectral density function,
\begin{equation}
   J_n(\omega) = \frac{\pi}{2}\sum_k \frac{g_{n,k}^2}{\omega_{n,k}}
                  \delta(\omega - \omega_{n,k})
                  \;.
\end{equation}
One of the great benefits of the path integral formalism is that it
places no restrictions upon the functional form of the spectral density
as opposed to many other approaches to open quantum systems.

Following our previous developments on the equilibrium reduced density
matrix,\cite{moix12} as well as those of several others on the 
full real time dynamics of the density matrix,\cite{stockburger02,zhou05} 
the nonlocality present in the influence functional can be substituted for
local interactions with stochastic auxiliary fields, 
which can then be efficiently sampled through Monte Carlo methods.
Formally, this is affected by applying a separate Hubbard-Stratonovich
transformation to each of the $N_{\rm A}$ terms in the influence functional.
Then Eq.~\ref{eq:if} can be exactly rewritten as 
\begin{align}
   F[\sigma]  =& \prod_{n=1}^{N_A} \int \mathcal D[\xi_n]\; w_n
      \exp\left(-\frac{1}{2\hbar}
      \int_0^t dt' \int_0^{t} d{t''} \xi_n(t') C_n^{-1}(t'-t'') \xi_n(t'')  +
      \frac{i}{\hbar}\int_0^t dt' V_n^{\rm A}(\sigma(t'))\xi_n(t') \right)
      \;,
\end{align}
where $w_n$ represents the normalization constant of the Gaussian functional
integral associated with the $n$-th bath.
The path integral involving the system variables is now completely local in
time and the auxiliary fields can be reinterpreted as a source of 
colored noise driving the system dynamics.
Thus individual samples of the absorption operator can be 
simply and straightforwardly calculated through the solution of a 
stochastic differential equation,
\begin{equation}
   \frac{d}{dt} \rho^{\rm A}(t) = -\frac{i}{\hbar} H^{\rm A}(t) \rho^{\rm A}(t) \;,
   \label{eq:absorption_prop}
\end{equation}
subject to the initial condition $\rho^{\rm A}(0)=I_{\rm s}$.
The stochastic Hamiltonian is given by,
\begin{equation}
   H^{\rm A}(t) =  H_{\rm s}^{\rm A} + \sum_n \xi_n(t) V_n^{\rm A}
   \;,
\end{equation}
and the {\it scalar}, complex-valued, Gaussian noise terms obey the 
correlations,
\begin{align}
   \left\langle \xi_n(t)\right \rangle  &= 0  \;, \nonumber \\ 
   \left\langle \xi_n(t)\xi_m(t')\right \rangle &= \delta_{nm} C_n(t-t')/\hbar
   \label{eq:noise} \;.
\end{align}
The exact time evolution of the absorption operator is obtained after
averaging the stochastic dynamics over realizations of the noise, 
$\mathbf I^{\rm A}(t) = \langle \rho^{\rm A}(t) \rangle_\xi$.
Here we have assumed that each of the baths is independent of all others.
To include correlated baths, one need only to replace the delta function
in Eq.~\ref{eq:noise} with the desired spatial correlations.
The generation of complex, Gaussian colored noise is discussed in 
Appendix~\ref{App:noise}.

The stochastic path integral equation for the absorption operator 
bears some similarity to the non-Markovian quantum state diffusion (NMQSD) 
approach recently proposed to compute the zero temperature absorption 
spectrum of excitonic systems.\cite{roden11}
While NMQSD is formally exact, all practical implementations to date 
have relied on approximations. 
In contrast, the present Eq.~\ref{eq:absorption_prop} is both formally and 
numerically exact, although it may prove fruitful to further explore the 
connections between the two approaches.

\subsection{Emission Operator}
\label{sec:em}

Due to the correlated initial state, the calculation of the 
emission operator is considerably more involved than that of the absorption.
The propagator in the path integral representation is,\cite{grabert88}
\begin{equation}
   U^{\rm D}(t,\hbar\beta) = \int {\mathcal D}[\sigma] 
   \int {\mathcal D}[\bar \sigma] 
   e^{-\frac i\hbar S_0^{\rm D}[\sigma] 
    - \frac 1\hbar S_0^{E,{\rm D}}[\bar \sigma] }
   F[\sigma,\bar \sigma]
   \;,
\end{equation}
where $S_0^{E,D}$ denotes the Euclidean action of the 
donor system Hamiltonian associated with the 
initial imaginary time propagation to the equilibrium state. 
The influence functional now contains three contributions from the 
respective propagations in real time, imaginary time, and the correlations
between the two, 
\begin{align}
   F[\sigma,\bar \sigma] =& \prod_n^{N_D} 
   \exp\left(-\frac{1}{\hbar} \int_0^t dt' \!\!\!
   \int_0^{t'} \!\!d{t''} \;
   V_n^{\rm D}(\sigma(t'))V_n^{\rm D}(\sigma(t'')) C_n(t'-t'') \right)
   \nonumber \\
   &\times
   \exp\left(\frac{1}{\hbar} \int_0^{\hbar \beta} d\tau' \!\!\!
   \int_0^{\tau'} \!\!d{\tau''} \;
   V_n^{\rm D}(\bar\sigma(\tau'))V_n^{\rm D}(\bar\sigma(\tau'')) 
   C_n(-i\tau'+i\tau'') 
   \right) \nonumber \\
   &\times
   \exp\left(\frac{i}{\hbar} \int_0^{\hbar \beta} d\tau' \!\!\!
   \int_0^{t} \!\!d{t'} \;
   V_n^{\rm D}(\bar \sigma(\tau'))V_n^{\rm D}(\sigma(t')) C_n^*(t-i\tau') 
   \right) 
   \;.
\end{align}
The bath correlation function is defined for complex arguments 
through the analytic continuation of Eq.~\ref{eq:ct} as,\cite{grabert88}
\begin{equation}
   C_n(z) = \frac1\pi \int_0^\infty d\omega \: J_n(\omega) 
   \frac{\cosh\left(\hbar\beta\omega/2 - i \omega z\right)}
        {\sinh\left(\hbar\beta\omega/2\right)}
        \;,\label{eq:Cz}
\end{equation}
where $z=t-i\tau$, and $0\le \tau\le \hbar\beta$.

As is readily seen the path integral expression for the emission operator
is considerably more complicated than the corresponding result for the
absorption operator.
Additionally, the coupling between the real and imaginary time paths in
the influence functional prevents a straightforward application of the 
Hubbard-Stratonovich transformation as was used previously for 
the absorption operator.
Fortunately, a simplification is possible.
The detailed balance relation in Eq.~\ref{eq:db_t} suggests that 
the emission operator may be computed in an identical manner to the 
absorption through the introduction of a complex time variable 
$z=t-i\hbar\beta$. 
Indeed, with this substitution in the path integral expressions above, 
the emission propagator may then be defined in 
an analogous fashion to Eq.~\ref{eq:abs_pi} except along a time-ordered 
contour in the complex time plane,\cite{grabert88}
\begin{equation}
   U^{\rm D}(z) = \int {\mathcal D}[\sigma] 
      e^{-\frac{i}{\hbar} S_0^{\rm D}[\sigma]} F[\sigma]\;.
\end{equation}
The influence functional similarly simplifies to,
\begin{equation}
   F[\sigma] = \prod_n^{N_D} \exp\left(-\frac{1}{\hbar} \int_0^z dz' \!\!\!
   \int\limits_{z'>z''} \!\!d{z''} \;
   V_n^{\rm D}(\sigma(z'))V_n^{\rm D}(\sigma(z'')) C_n(z'-z'') \right)
   \;.
\end{equation}

With these results, the Hubbard-Stratonovich transformations and 
stochastic Schr{\"o}dinger equation are then formally equivalent to those 
of the absorption operator presented in Sec.~\ref{sec:abs}. 
The complex time evolution of samples of the emission operator obeys
the equation
\begin{equation}
   \frac{d}{dz} \rho^{\rm D}(z) = -\frac i\hbar H^{\rm D}(z) \rho^{\rm D}(z)
   \;, \label{eq:emission_prop}
\end{equation}
where the integration proceeds from time $z=0$ to the complex time 
$z=t-i\hbar\beta$ subject to the initial condition, 
$\rho^{\rm D}(0)=I_{\rm s}$.
Similarly, the stochastic complex time-dependent Hamiltonian is given by
\begin{equation}
   H^{\rm D}(z) = H_{\rm s}^{\rm D} + \sum_n \xi_n(z) V_n^{\rm D} \;,
\end{equation}
where the scalar noise components are again of zero mean and correlation:
\begin{align}
   \left\langle \xi_n(z) \right\rangle & = 0  \;, \nonumber \\
   \left\langle \xi_n(z) \xi_m(z) \right\rangle 
      & = \delta_{nm} C_n(z-z')/\hbar \;.
\end{align}
As with the absorption, the numerically exact emission operator is obtained 
after the stochastic average over the noise variables,
$\mathbf E^{\rm D}(t)^* = \langle \rho^{\rm D}(t-i\hbar\beta) \rangle_\xi
/\langle Z\rangle_\xi$ where $Z={\rm Tr}\left[\rho^{\rm D}(-i\hbar\beta)\right]$.

\begin{figure}
   \includegraphics*[width=0.55\textwidth]{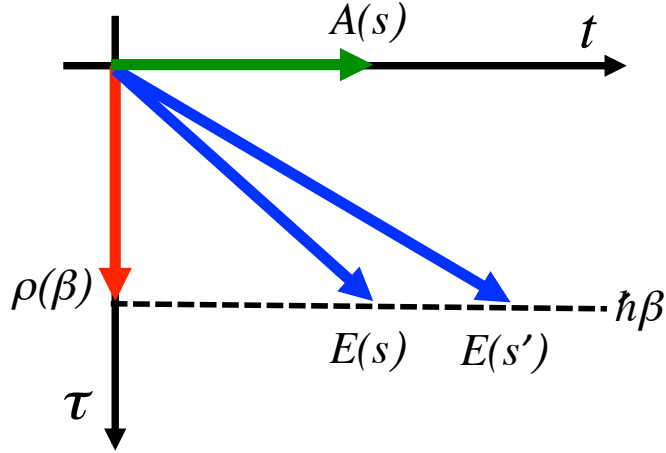}
   \caption{
      The integration contours in the complex time plane $z=t-i\tau$ 
      used in the various calculations.
      Integrating along $z=t$ (green arrow) yields the absorption operator 
      while the contour $z=-i\tau$ (red arrow) results in the equilibrium
      reduced matrix.
      The blue arrows characterize the independent contours needed to 
      generate the emission operator at times $s$ and $s'$.
   }\label{fig:contours}
\end{figure}

It should be noted that the stochastic path integral expression in 
Eq.~\ref{eq:emission_prop} represents a generalized form of the 
stochastic propagation. 
It contains two interesting limits that are represented schematically 
in Fig.~\ref{fig:contours}.
For example, if the imaginary time component of the propagation
is set to zero such that $z=t$, then one immediately recovers the absorption 
operator obtained above in Eq.~\ref{eq:absorption_prop}.
Alternatively, if the real time component of the integration contour 
is set to zero, then one recovers the pure imaginary time evolution of the
equilibrium reduced density matrix propagation explored in our earlier
work.\cite{moix12}
This leads to the interesting result that the emission operator at 
$t=0$ is simply the exact equilibrium reduced density matrix.

\subsection{Computational considerations}

There are several points with regards to the stochastic formulation 
that should be emphasized.
Firstly, a generalized stochastic approach to compute the real time dynamics 
of the entire reduced density matrix has recently been 
explored.\cite{cao96,stockburger02,zhou05}
In that case, the presence of complex noise generally leads to very slow 
convergence of the stochastic average as the length of the simulation
increases.
The approach presented here, and in particular Eq.~\ref{eq:absorption_prop}, 
represents a simplified version of those works,
and thus directly inherits their numerical difficulties.
However, the redeeming feature of the present approach 
is that the decay time of the absorption and emission correlation functions 
is much shorter than the corresponding relaxation time of the pure 
real time dynamics.
As a result, brute force convergence of the stochastic path integral 
is generally possible with a reasonable number of Monte Carlo 
samples ($10^5 - 10^6$), although the low temperature regime
can be more demanding.

Secondly, there is a subtle difficulty that should be discussed with regards 
to the calculation of the emission spectrum.
As seen from Fig.~\ref{fig:contours}, the propagations to times 
$s-i\hbar\beta$ and $s'-i\hbar\beta$ evolve along different contours 
in the complex time plane. 
The bath correlation functions evaluated along these two contours are
different,
and thus the emission operators $\mathbf E(s)$ and $\mathbf E(s')$ 
require completely independent calculations.
In principle, this should increase the cost of computing the emission spectrum
by a factor of the number of time steps with respect to that of the absorption.
However, this is not the case since the presence of the 
imaginary time component in the propagation greatly improves the convergence
properties of the Monte Carlo calculation.\cite{berne86,cao96}
While the computational cost of the emission spectrum is more expensive,
it is not prohibitive.

Finally, the inclusion of static disorder in the absorption operator 
calculation is trivial, but less so for that of the emission operator.
In the former case, the averages over the noise and disorder commute, 
and thus the two may be computed simultaneously.
That is, the disorder-averaged absorption spectrum should incur practically no
additional computational cost over that of the bare absorption spectrum.
However, the presence of the partition function in the denominator of the
emission operator demands that the average over the disorder must
be computed independently of the average over the noise. 
As a result the disorder-averaged emission spectrum, although straightforward,
may be quite costly from a computational perspective.

\section{Numerical Results}\label{sec:results}

Although the path integral formalism is valid for any spectral density,
below we will focus on the standard Drude-Lorentz form so that 
benchmark results from the HEOM formalism can be obtained. 
In a forthcoming work, we will examine the influence of the spectral density
on the spectra and energy transfer rates of the light harvesting system 
LH2.\cite{moix14b}
The Drude spectrum is defined by
\begin{align}
   J(\omega) & = 2\lambda\frac{\omega\gamma}{\omega^2 + \gamma^2} \;, 
   \label{eq:spec_dens}
\end{align}
where $\gamma$ is the cutoff frequency and the reorganization energy $\lambda$
is defined such that,
\begin{equation}
   \lambda = \frac{1}{\pi}\int_0^\infty d\omega\; \frac{J(\omega)}{\omega}
   \;.
\end{equation}
As is commonly assumed, we take the spectral densities for each of the 
independent baths to be equivalent and, unless otherwise specified, 
fix the reorganization energy at $\lambda=200$ cm$^{-1}$ and the cutoff 
frequency to $\gamma=53$ cm$^{-1}$ (10 ps$^{-1}$).

\subsection{Emission spectra}

Before presenting the MCFT rates, we will first focus on the 
far-field spectra. 
The absorption and emission spectrum can be computed by combining the
knowledge of the corresponding operators defined above
in Eqns.~\ref{eq:It} and~\ref{eq:Et} with the respective transition dipole 
moment vectors of the chromophores ($\vec \mu$), by
\begin{align}
   I_i^{\rm A}(\omega) & = \int_{-\infty}^\infty dt \; e^{+i\omega t} 
      \sum_{n,m} 
      \left(\vec\varepsilon_i \cdot \vec \mu_m^{\rm A}\right)
      \left(\vec\varepsilon_i \cdot \vec \mu_n^{\rm A}\right) 
      \mathbf I_{mn}^{\rm A}(t) 
      \label{eq:absorption_ff} \\
   E_i^{\rm D}(\omega) & = \int_{-\infty}^\infty dt \; e^{-i\omega t} 
      \sum_{n,m} 
      \left(\vec\varepsilon_i \cdot \vec \mu_m^{\rm D}\right)
      \left(\vec\varepsilon_i \cdot \vec \mu_n^{\rm D}\right) 
      \mathbf E_{mn}^{\rm D}(t) 
      \label{eq:emission_ff}
\end{align}
where $\vec \varepsilon_i$ is a unit vector characterizing the
polarization of the incident radiation field that projects onto the
dipole moment vector of each chromophore.
As noted above, the path integral evaluation of the emission spectrum contains 
the absorption spectrum as a limiting case, and hence we will focus only on
the former here.

\begin{figure}
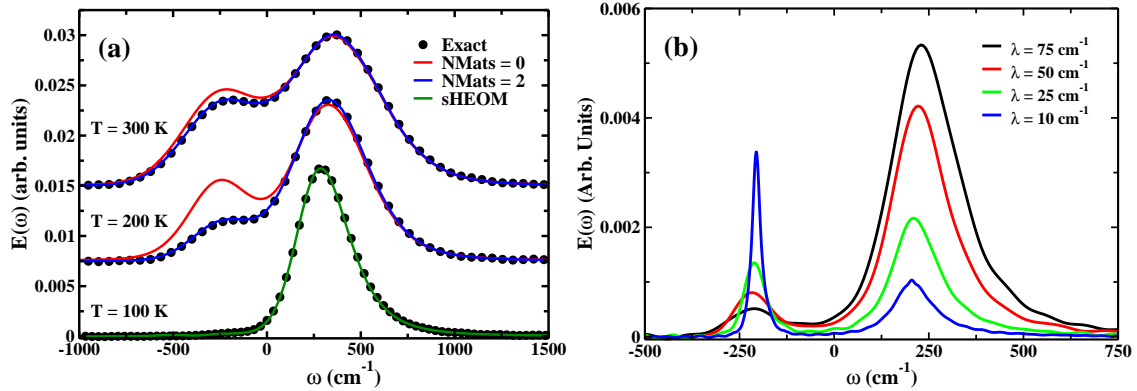

   \includegraphics[width=0.45\textwidth]{./figure2a.eps}
   \includegraphics[width=0.45\textwidth]{./figure2b.eps}
   \caption{
      (a) The emission spectrum at $T=100$, $200$, and $300$ K
      in a model two level system comparing the present stochastic 
      path integral calculations with the corresponding
      HEOM results with $0$ (red) and $2$ (blue) Matsubara terms.
      The number of hierarchy tiers required for convergence in each case
      is $12$.
      The bath is defined by the reorganization energy 
      $\lambda=200$ cm$^{-1}$, and cutoff frequency, $\gamma=53$ cm$^{-1}$.
      For clarity, the results at $T=200$, and $300$ K are vertically 
      offset by $0.0075$, and $0.015$, respectively.
      At $T=100$ K we cannot converge the standard HEOM and only the 
      stochastic HEOM (green) formalism can produce reliable results.
      (b) The reorganization energy dependence of the emission spectrum
      at $T=100$ K.
      The remaining parameters are the same as in (a).
   }
   \label{fig:2ls_spec}
\end{figure}

As a preliminary benchmark calculation to prove the efficacy
of the path integral approach, Fig.~\ref{fig:2ls_spec} displays
the stochastic path integral results for the emission spectrum
in the unbiased two level system that was explored in papers I and II.
The system Hamiltonian $H_{\rm s}=V\sigma_x$ with $V=200$ cm$^{-1}$,
leads to highly delocalized exciton states, and thus
serves as an interesting test case to assess the validity of the
MCFT formalism as well as that of approximate perturbative approaches.
For simplicity, the dipole moment operators have been chosen to be equivalent
for each site and in each direction such that $\vec \mu_m=1$ 
in Eqns.~\ref{eq:absorption_ff} and~\ref{eq:emission_ff}.
Because of this choice, the spectra are determined from the simple sum 
of all of the elements of the emission operator (cf.~Eq.~\ref{eq:emission_ff}).
In Fig.~\ref{fig:2ls_spec}, the path integral results at $200$ 
and $300$ K are compared with the corresponding results from the 
standard HEOM approach shown with increasing number of Matsubara terms.
The HEOM results are seen to eventually converge to the path integral results,
although even for this relatively simple two level system, 
the hierarchy results are difficult to converge and require both a 
large number of hierarchy tiers as well as several Matsubara terms.
At the lowest temperature of $T=100$ K shown in Fig.~\ref{fig:2ls_spec}, 
the standard hierarchy calculations can not be converged with respect 
to the number of Matsubara terms,
and the sHEOM approach must be used to generate the 
numerically exact results.\cite{moix13a,zhu13}
Note that at each temperature, the hierarchy results and present path integral 
results are in precise agreement.
However, compared with the hierarchy calculations, 
the stochastic approach developed here is more straightforward
both in terms of implementation and convergence.
Additionally, since the stochastic formalism is a Monte Carlo method, 
it is trivially parallelized and free from 
the memory demands that plague other density matrix approaches such 
as the HEOM or QUAPI.
In the case that further improvements to the computational efficiency 
of the emission path integral are necessary, 
a very useful and accurate approximation can be employed 
which is discussed in Appendix~\ref{App:emission}.

As is readily seen, the spectra in Fig.~\ref{fig:2ls_spec}a are comprised of 
two peaks centered around the eigenstates of the total system Hamiltonian.
While the intensity of the peak at positive frequencies is nearly independent
of temperature, that of the low energy peak steadily decreases and 
vanishes at the lowest temperature shown of $T=100$ K.
This is in stark contrast to the behavior expected from an
isolated two level system where the emission spectrum 
can be computed analytically as
\begin{align}
   E(\omega) &= \sum_{i=1}^2 P_i 
                \left(1+\frac{V}{\epsilon_i}\right)
                \delta(\omega-\epsilon_i)
                \;, \label{eq:isolated_emission}
\end{align}
with the eigenstate energies, $\epsilon_{1,2}=\pm \sqrt{V^2+\Delta^2}$,
$V$ is the electronic coupling, $\Delta$ is the bias,
and the eigenstate population, $P_i=e^{-\beta \epsilon_i}/Z$.
As seen from Eq.~\ref{eq:isolated_emission} the spectra are composed 
of two peaks centered at the eigenfrequencies of the system with 
intensities that are weighted by the respective Boltzmann populations 
of the two states.
At low temperature the population localizes in the ground state, and the
spectrum shifts to the red.
In Fig.~\ref{fig:2ls_spec}, the opposite occurs and 
a blue shift is clearly seen with decreasing temperature.
This behavior is a result of the strong system bath coupling.
To demonstrate this effect more clearly, Fig.~\ref{fig:2ls_spec}b 
displays the reorganization energy dependence of the emission spectra 
at the lowest temperature shown in Fig.~\ref{fig:2ls_spec}a of $T=100$ K.
Only at very weak coupling, does the spectra resemble that expected
for the isolated system from Eq.~\ref{eq:isolated_emission},
with the emission dominated by the low energy eigenstate of the system.
However, as the coupling increases in Fig.~\ref{fig:2ls_spec}b, 
the weighting of the two peaks is redistributed towards the higher lying 
eigenstate resulting in a steady shift to the blue.
As discussed in papers I and II, the equilibrium state of the total system and 
bath cannot be written in a factorized form as in 
Eq.~\ref{eq:isolated_emission}, particularly when the
temperature is low and the system-bath coupling large, as is the case here. 
This is the key feature that is responsible for the drastic failure of 
standard perturbative approximations to the emission spectra as well as 
the counterintuitive temperature dependence seen in Fig.~\ref{fig:2ls_spec}a.

\subsection{MCFT rates}

\begin{figure}
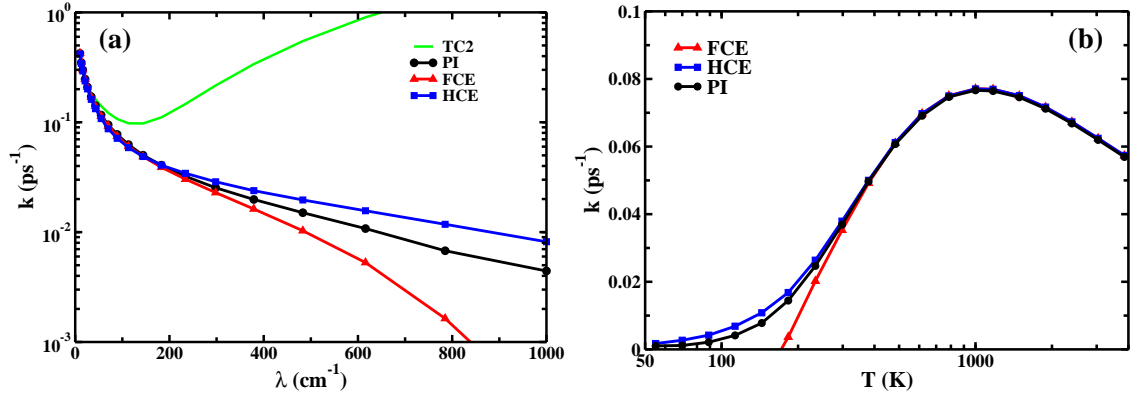

   \includegraphics*[width=0.45\textwidth]{./figure3a.eps}
   \includegraphics*[width=0.45\textwidth]{./figure3b.eps}
   \caption{
      (a)
      Energy transfer rates between two symmetric two level systems 
      as a function of the reorganization energy.
      The cutoff frequency, $\gamma=53$ cm$^{-1}$ and the temperature
      is $T=300$ K.
      The black dots, green lines, red triangles, and blue squares 
      denote the results 
      from the present stochastic path integral approach, 
      the second order time-convolution (TC2) master equation,\cite{jang04}
      the full cumulant expansion (FCE),\cite{ma14a,banchi13} 
      and the hybrid cumulant expansion (HCE),\cite{ma14b} respectively.
      (b)
      The energy transfer rates as a function of the temperature.
      The reorganization energy is $\lambda=200$ cm$^{-1}$ and 
      the cutoff frequency, $\gamma=53$ cm$^{-1}$.
      The results from the TC2 approach lie outside the scale of the graph
      for all temperatures shown.
   }\label{fig:tls_rates_l}
\end{figure}

We next consider the multi-chromophoric energy transfer,
where both the donor and acceptor complexes are comprised of the symmetric two
level system analyzed in Fig.~\ref{fig:2ls_spec}.
That is, each complex is described by the system Hamiltonian, 
$H_{\rm s}=V\sigma_x$, 
and the donor-acceptor couplings (Eq.~\ref{eq:h_da}) are set to the 
constant value, $J_{nm}^{\rm DA}=10$ cm$^{-1}$.
This weak coupling ensures that the perturbative MCFT formalism is valid
and is also characteristic of many natural systems.\cite{ma14a}
The energy transfer rates computed as a function of the system-bath coupling 
strength are displayed in Fig.~\ref{fig:tls_rates_l}a.
Although not shown, the rates from the HEOM formalism are in precise 
agreement with the present path integral results 
in the region where the former can be converged
(up to $\lambda=600$ cm$^{-1}$).
As the transfer occurs between two symmetric systems, the transfer 
rates are monotonically decreasing functions of the system-bath
coupling strength as would be expected from a simple analysis
based on the standard F{\"o}rster theory.

Also included in Fig.~\ref{fig:tls_rates_l}a, is a comparison of the exact
energy transfer rates with many of the commonly used perturbative methods.
The TC2 is the standard second order, time-convolution master equation
previously explored.\cite{jang04}
As it is based upon the approximation of weak system-bath coupling,
its validity is rather limited, and is generally inapplicable
to many interesting physical systems, such as light-harvesting complexes,
where the system-environment couplings can not be considered as small.
The full cumulant approximation (FCE) explored in paper I 
and Ref.~\onlinecite{banchi13} provides reliable results 
over a much larger region of the parameter space as compared with the TC2,
although it too begins to break down at very large system-bath couplings
and eventually produces unphysical negative rates.
The failure of both the TC2 and the FCE lies in their inaccurate 
treatment of the correlated initial state.
Clearly, a perturbative expansion around a factorized initial state
is qualitatively incorrect at large-system bath coupling.
In order to overcome this difficulty, paper II explored 
an expansion around the numerically exact equilibrium reduced density 
matrix, which can be straightforwardly 
obtained  through imaginary time path integral 
techniques.\cite{moix12}
As seen in Fig.~\ref{fig:tls_rates_l}a this HCE technique provides 
a uniformly reliable approximation to the energy
transfer rate, even at very strong system bath couplings.

Fig.~\ref{fig:tls_rates_l}b displays the temperature dependence of 
the MCFT rates.
Qualitatively, the results follow the predictions of Marcus theory
displaying a maximum as a function of temperature.
However, the Marcus rate formula predicts a maximum at 
$2\lambda/k_{\rm B}\approx  600$ K
which is considerably lower than that observed from the exact calculations.
Additionally, Marcus theory predicts that the
energy transfer rate should vanish as the temperature decreases to zero.
This is clearly not borne out in the exact results as the MCFT
rates decrease to a finite value at low temperature due to non-vanishing 
quantum fluctuations.
As with the system-bath coupling dependence, the approximate
perturbative results are also included. 
It is seen that below room temperature, the accuracy of the FCE 
quickly degrades and eventually produces negative energy transfer 
rates. 
The results from the TC2 method are outside the scale of the 
graph at all temperatures, which is not entirely unexpected as the
system-bath coupling strength here is of comparable magnitude to all the other
system parameters.
However, the hybrid method provides reliable results across the entire 
parameter range, and also captures the plateau in the rates at 
low temperature.
In a forthcoming work, it is demonstrated that the HCE is
capable of capturing the temperature and system-bath coupling dependence 
of the energy transfer rates between two LH2 complexes while the 
failure of the TC2 and FCE is even more dramatic than seen here.\cite{moix14b}

\section{Conclusions}

A stochastic path integral approach to compute the energy transfer 
rates between weakly coupled multi-chromophoric complexes has been presented.
As a consequence of the MCFT formalism, one also has 
immediate access to the exact steady-state far-field absorption and emission
spectra of the respective donor and acceptor complexes.
The calculations of the absorption and emission operators 
require only the straightforward numerical solution
of a stochastic differential equation, and the only difficulty lies
in the convergence of the Monte Carlo average.
As opposed to many other numerically exact approaches, 
the method developed here is amenable to any form of the
spectral density, and can readily treat the low temperature and strong
coupling regimes.
To our knowledge, the present path integral approach is the
only method currently available that can accommodate such a broad range of
system parameters in relatively large excitonic systems.

The numerical results presented here provide a systematic
analysis of the role of the temperature and system-bath coupling strength
on the emission spectra and energy transfer rates in model multi-chromophoric
systems.
As seen in Fig.~\ref{fig:tls_rates_l} the exact MCFT rates 
serve as a stringent benchmark for approximate analytic methods.
Whereas the standard perturbative approaches often yield qualitatively 
incorrect results, the hybrid cumulant expansion (HCE) technique developed 
in paper II\cite{ma14b} can provide uniformly reliable results for the energy 
transfer rates across a large range of the physically accessible parameter
space.

\section{Acknowledgments}

This work was supported by the NSF (Grant No. CHE-1112825).
J.~Moix and J.~Ma have been supported by the Center for Excitonics,
an Energy Frontier Research Center funded by the US Department of Energy,
Office of Science, Office of Basic Energy Sciences under Award
No.~DE-SC0001088.
This work used the Extreme Science and Engineering Discovery Environment
(XSEDE), which is supported by National Science Foundation grant number
OCI-1053575.

\appendix

\section{Noise sampling}\label{App:noise}

Sampling the complex noise required for the absorption and emission operators
is not completely trivial.
The main difficulty is that the bath correlation function must be
reproduced by $\langle \xi(t)\xi(t')\rangle$ rather than the Hermitian form
$\langle \xi^*(t) \xi(t')\rangle$.
To proceed, the correlation function can be split into its real and imaginary 
components,
\begin{equation}
   C(t) = C_r(t) + iC_i(t)
\end{equation}
and the influence functional rewritten as
\begin{equation}
   F[\sigma] = \exp\left[-\frac{1}{2\hbar}
          \int_0^t dt' \int_0^{t} d{t''} V(\sigma(t'))V(\sigma(t'')) 
          \left[C_r(t'-t'') + iC_i(|t'-t''|)\right] \right] \;.
\end{equation}
Hubbard-Stratonovich transformations are then applied to each term
separately, leading to
\begin{align}
   F[\sigma]  =& \int \mathcal D[\zeta]\; w_\zeta
      \exp\left[-\frac{1}{2\hbar}
      \int_0^t dt' \int_0^{t} d{t''} \zeta(t') C_r^{-1}(t'-t'') \zeta(t'')  +
      \frac{i}{\hbar}\int_0^t dt' V(\sigma(t'))\zeta(t') \right]\nonumber \\
   \times & \int \mathcal D[\nu] \; w_\nu
      \exp\left[-\frac{1}{2\hbar}
      \int_0^t dt' \int_0^{t} d{t''} \nu(t') C_i^{-1}(|t'-t''|) \nu(t'')  +
      \frac{1-i}{\sqrt{2}\hbar}\int_0^t dt' V(\sigma(t'))\nu(t') \right]\;,
\end{align}
where $w_\zeta$ and $w_\nu$ denote the respective normalization constants.
Thus the noise characteristics are
\begin{align}
   \left\langle \zeta(t)\right \rangle  &= 0  
   &\left\langle \nu(t)\right \rangle  &= 0  \nonumber \\ 
   \left\langle \zeta(t)\zeta(t')\right \rangle &= C_r(t-t')  
   & \left\langle \nu(t)\nu(t')\right \rangle &= C_i(|t-t'|) \;,
\end{align}
and the autocorrelation function of the combined 
process, $\xi(t)=\zeta(t) + \sqrt{i}\nu(t)$ is readily seen 
to reproduce the desired bath correlation function, $C(t)$.

Numerically sampling the real noise governed by the correlation, $C_r(t)$, 
is straightforward since this kernel is strictly positive semi-definite.
Sampling such noise has been discussed in detail in our previous
works.\cite{moix12,moix13a}
One simply filters white noise with a kernel computed from
the Cholesky decomposition of the Toeplitz matrix constructed from $C_r(t)$.

Sampling the noise for the imaginary part of the correlation function 
is less straightforward. 
The kernel, $C_i(t)$ is not positive definite since $C_i(0)=0$, 
so that the Cholesky decomposition approach is not applicable.
To cope with this, we have employed the approach suggested in 
Refs.~\onlinecite{zhou08,cao96}.
First an eigen decomposition of the correlation matrix, 
$C_{i,nm} = C_i(|t_n-t_m|)$, is performed,
and the diagonal eigenvalue matrix is sorted into a nonnegative 
($\boldsymbol \Lambda^+$) matrix and the remainder ($\boldsymbol \Lambda^-$),
which are both of the same dimension as $\mathbf C_i$, such that
\begin{equation}
   \mathbf C_i= \mathbf U\left[\boldsymbol \Lambda^+ 
                            + \boldsymbol \Lambda^-\right]\mathbf U^T
                            \;.
\end{equation}
The positive components are sampled in the usual fashion by filtering 
the appropriate kernel with white noise, while 
the negative components are sampled by taking the absolute value of the 
negative eigenvalues followed by a rotation with the complex unit. 
The desired noise sequence is then given by
\begin{equation}
   \vec \nu = \mathbf U^T\left[\left(\boldsymbol\Lambda^+\right)^{1/2}
      +  i|\boldsymbol\Lambda^-|^{1/2}\right] \vec \Omega \;,
\end{equation}
where $\vec \Omega$ represents a realization of independent white noise 
terms.
Using the properties of white noise, it is readily seen that the 
autocorrelation of $\nu(t)$ faithfully reproduces
the desired imaginary part of the bath correlation function.

\section{Approximate Emission}\label{App:emission}

\begin{figure}
   \includegraphics[width=0.6\textwidth]{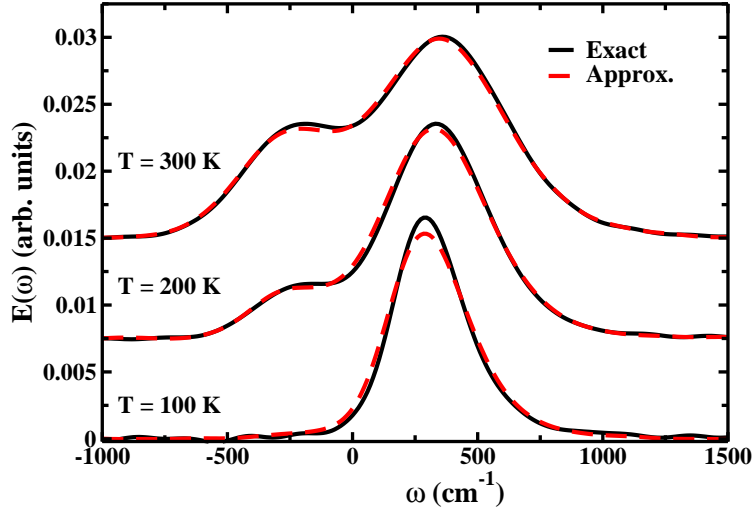}
   \caption{
      The exact two-level system emission spectra reproduced from
      Fig.~\ref{fig:2ls_spec} (solid black) compared with the 
      approximate emission spectra (dashed red) computed by ignoring 
      the imaginary part of the bath correlation function.
      The parameters are identical to those in Fig.~\ref{fig:2ls_spec}.
   }\label{fig:tls_approx_emission}
\end{figure}

A very accurate approximation to the emission operator can be made by 
simply ignoring the imaginary part of the bath correlation function 
in Eq.~\ref{eq:Cz}. 
This simplification generally reduces the number of Monte Carlo 
samples required to converge the stochastic path integral by at least 
an order of magnitude.
For the purely real-time dynamics of the absorption operator, 
ignoring $C_i(t)$ leads to an extended Haken-Strobl model which 
rarely provides satisfactory results.
However, for emission operator, the real-time and imaginary-time dynamics
are intertwined so that the analysis is more subtle. 
In this case, there are still non-unitary contributions to the dynamics even 
if the Hamiltonian is purely real due to the complex-time evolution.
To better understand this seemingly drastic approximation, it is
useful to analyze the complex-time bath correlation function.
It is readily seen that the bath correlation function evaluated 
along the imaginary time axis to $z=-i\hbar\beta$ is a purely real 
quantity for {\it any} spectral density.
This case corresponds to the equilibrium reduced density matrix which is 
a purely real quantity if the Hamiltonian is real.
Thus for small real times --during which time the emission operator has often
substantially decayed-- the imaginary part of the correlation function 
is also negligibly small.
This approximation is particularly accurate in the high temperature limit 
where the increasingly broad emission spectra are a result of the 
increasingly rapid decay of the emission operator.
In summary, in short-time limit ignoring $C_i(z)$ is a reasonable 
approximation.
In Fig.~\ref{fig:tls_approx_emission} the exact results for the emission
spectrum of the two level system are reproduced from 
Fig.~\ref{fig:2ls_spec} along with the corresponding 
results from the approximation scheme discussed here where the imaginary part
of the complex-time bath correlation function has been set to zero. 
Only at the lowest temperature of $T=100$ K are there any significant
differences between the exact and approximate emission spectra.
In fact, comparison with Fig.~\ref{fig:2ls_spec} indicates that even at 
$T=300$ K, the approximate emission spectrum is more accurate than the 
HEOM results computed without including Matsubara terms.

%
\end{document}